\documentclass{article}
\usepackage{spconf,amsmath,graphicx}
\usepackage{hyperref}
\usepackage{enumitem}
\usepackage{booktabs}
\usepackage{array}    
\usepackage{xcolor}
\setlist{nosep, leftmargin=14pt}
\usepackage{caption}
\usepackage{mwe} 
\usepackage{etoolbox}  
\usepackage{hyperref}
\makeatletter
\renewcommand{\section}{\@startsection{section}{1}{0pt}{-1.5ex plus -1.3ex minus -1ex}{1.3ex plus .2ex}{\normalfont\Large\bfseries}}
\renewcommand{\subsection}{\@startsection{subsection}{2}{0pt}{-1.3ex plus -1.3ex minus -1ex}{1ex plus .2ex}{\normalfont\large\bfseries}}

\apptocmd{\thebibliography}{\setlength{\itemsep}{0.05ex}}{}{}

\title{CSF-Net: Cross-Modal Spatiotemporal Fusion Network for Pulmonary Nodule Malignancy Predicting}
\name{
\begin{tabular}[t]{c}
Yin Shen$^1$,
Zhaojie Fang$^1$,
Ke Zhuang$^1$,
Guanyu Zhou$^1$,
Xiao Yu$^1$,
Yucheng Zhao$^2$,\\
Yuan Tian$^5$,
Ruiquan Ge$^{1, 4,*}$,
Changmiao Wang$^{3,*}$,
Xiaopeng Fan$^4$,
Ahmed Elazab$^6$
\thanks{$\!\!\!^*$Correspondings : gespring@hdu.edu.cn, cmwangalbert@gmail.com}
\end{tabular}}
\address{\fontsize{11pt}{5pt}\selectfont$^1$Hangzhou Dianzi University, Hangzhou, China \quad
\fontsize{11pt}{5pt}\selectfont$^2$Tandon School of Engineering, New York University, New York, USA\\
\fontsize{11pt}{5pt}\selectfont$^3$Shenzhen Research Institute of Big Data, Shenzhen, China $\,$
\fontsize{11pt}{5pt}\selectfont$^4$Hangzhou Institute of Advanced Technology, China \\
\fontsize{11pt}{5pt}\selectfont$^5$The Second Affiliated Hospital, School of Medicine, The Chinese University of Hong Kong, Shenzhen, China \\
\fontsize{11pt}{5pt}\selectfont$^6$Shenzhen University, Shenzhen, China
    }
\begin{document}

%
\maketitle
\begin{abstract}
Pulmonary nodules are an early sign of lung cancer, and detecting them early is vital for improving patient survival rates. Most current methods use only single Computed Tomography (CT) images to assess nodule malignancy. However, doctors typically make a comprehensive assessment in clinical practice by integrating follow-up CT scans with clinical data. To enhance this process, our study introduces a Cross-Modal Spatiotemporal Fusion Network, named CSF-Net, designed to predict the malignancy of pulmonary nodules using follow-up CT scans. This approach simulates the decision-making process of clinicians who combine follow-up imaging with clinical information. CSF-Net comprises three key components: spatial feature extraction module, temporal residual fusion module, and cross-modal attention fusion module. Together, these modules enable precise predictions of nodule malignancy. Additionally, we utilized the publicly available NLST dataset to screen and annotate the specific locations of pulmonary nodules and created a new dataset named NLST-cmst. Our experimental results on the NLST-cmst dataset demonstrate significant performance improvements, with an accuracy of 0.8974, a precision of 0.8235, an F1 score of 0.8750, an AUC of 0.9389, and a recall of 0.9333. These findings indicate that our multimodal spatiotemporal fusion approach, which combines follow-up data with clinical information, surpasses existing methods, underscoring its effectiveness in predicting nodule malignancy. Our code can be accessed at \href{https://github.com/hdusyz/CSF-NET}{https://github.com/hdusyz/CSF-NET}.
\end{abstract}
\section{Introduction}
\label{sec:intro}
Lung cancer, originating from lung cells, is a rapidly growing malignancy that poses a serious global public health threat. According to the World Health Organization and the American Cancer Society, there were approximately 2.5 million new cases of lung cancer worldwide in 2022, accounting for 12.4\% of new cancer diagnoses, and around 1.8 million lung cancer-related deaths, representing 18.7\% of all cancer deaths \cite{bray2024global}. As common lesions, pulmonary nodules may progress to lung cancer if not closely monitored, making follow-up management crucial. The National Lung Screening Trial (NLST) \cite{national2011national} in the United States has shown that regular imaging and lung function assessments can track nodule changes, predicting malignancy risks. Follow-ups help in early detection, increasing survival rates. However, current assessments largely rely on single CT scans, limiting insights into nodule progression and potentially leading to late diagnoses. Regular imaging follow-ups enable dynamic data collection on nodule changes, improving malignancy prediction and prevention for better patient health outcomes.

In pulmonary nodule classification and prediction research, significant progress has been made in applying cross-temporal classification networks. For instance, Jiang \textit{et al.} \cite{jiang2021learning} proposed an efficient classification method based on spatial information from CT images, while Liao \textit{et al.} \cite{8642524} employed a Leaky Noisy OR network for 3D deep learning to enhance understanding of the temporal dimension, although they did not fully explore temporal variations. Recently, spatiotemporal network research has begun to integrate multi-timepoint CT image data to improve classification and prediction accuracy. For example, Liu \textit{et al.}'s LSTM model \cite{liu2022study} has made advancements in capturing disease progression. Veasey \textit{et al.} \cite{9195164} and Khademi \textit{et al.} \cite{khademi2023spatio} further addressed the issues of interpretability and computational inefficiency in previous models, achieving better cancer malignancy predictions by balancing spatial details and temporal dynamics using convolutional autoencoders combined with Swin Transformers and Siamese convolutional attention networks. Additionally, multimodal clinical data has shown great potential in enhancing predictive accuracy. Structural features from CT images, combined with scales and histological slides, can further improve predictions. Guo \textit{et al.} \cite{10635747} proposed the PE-MVCNET to effectively utilize multi-view and cross-modal fusion for pulmonary embolism prediction. Yu \textit{et al.} proposed ICH-PRNet \cite{yu2025prnet}, using cross-modal fusion of text and imaging for accurate intracerebral hemorrhage prognosis classification. Likewise, multimodal fusion techniques have been successfully applied in cancer subtype classification \cite{10539123}. Furthermore, integrating imaging with genomic data has shown effectiveness in predicting cancer recurrence and survival rates \cite{subramanian2020multimodal,ellen2023autoencoder}. Aslani \textit{et al.} \cite{aslani2024enhancing} demonstrated the potential of multimodal and multi-timepoint fusion in predicting lung nodule malignancy through their DeepCAD method.

Despite the notable achievements of previous studies, several limitations remain. First, these studies often overlook the significance of integrating clinical data in multi-timepoint predictions, missing the advantages of multimodal fusion. Second, they typically use simple concatenation across the temporal dimension when combining multimodal and multi-timepoint data, failing to fully utilize the rich temporal features. To address these limitations, we introduce CSF-Net, a novel approach for predicting pulmonary nodule malignancy. Our method offers the following key contributions:

\begin{itemize}
    \item We integrated multi-timepoint follow-up data with clinical data to enhance the prediction of future malignancy of pulmonary nodules.
    \item We designed a temporal residual fusion module that effectively combines features across different timepoints.
    \item We employed cross-modal attention fusion to learn spatiotemporal and clinical features, thereby improving the model's accuracy and robustness.
\end{itemize}

\begin{figure*}[t!]
    \begin{center}
    \includegraphics[width=0.65\textwidth,height=0.3\textheight]{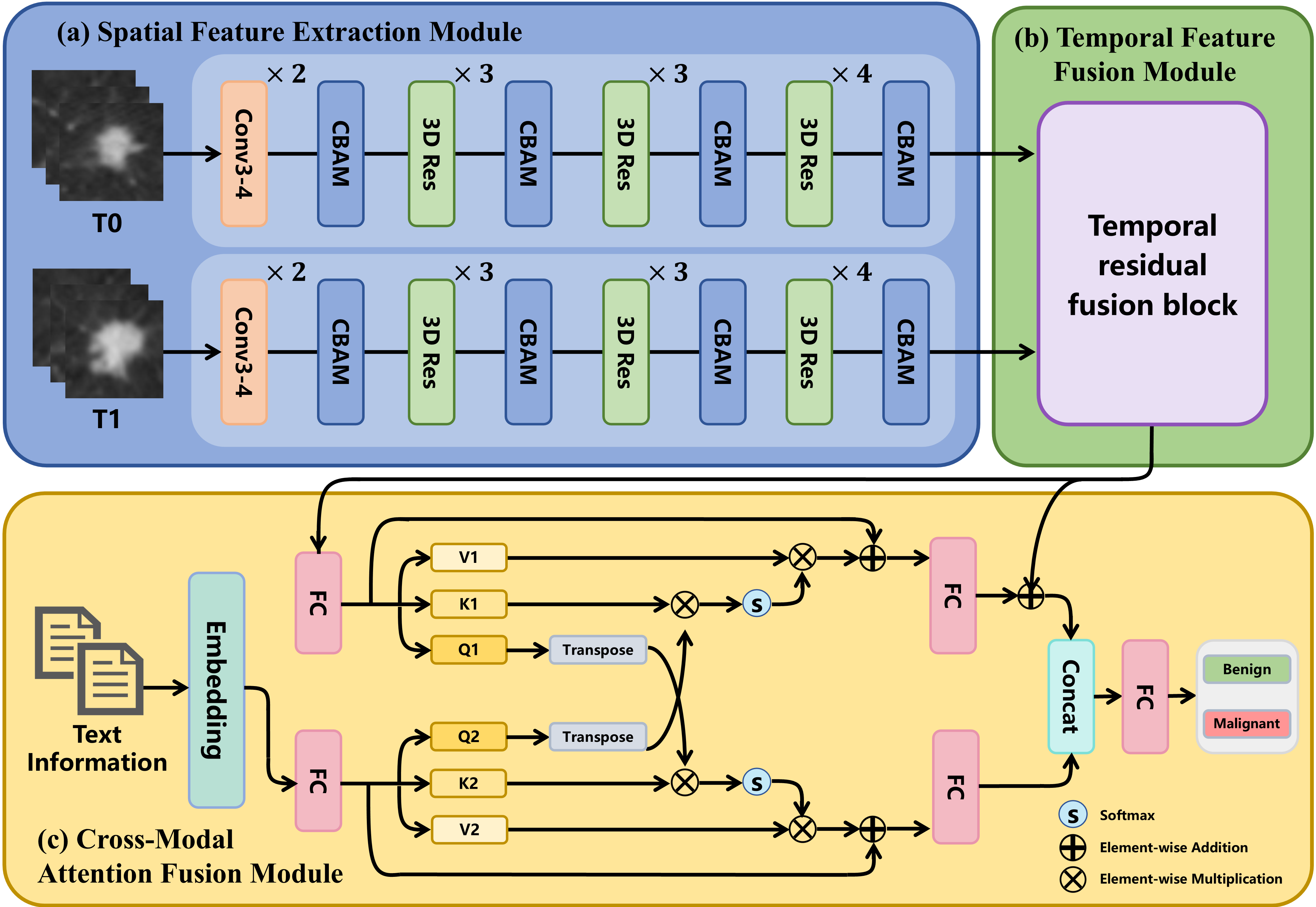}
    \end{center}
   \caption{The proposed architecture for CSF-Net. It consists of three components: (a) spatial feature extraction module, (b) temporal residual fusion module, and (c) CMAF module. The spatial module combines ResNet with the CBAM for enhanced feature representation. The temporal residual fusion module integrates features from different time points to capture correlations. The CMAF module uses cross-modal attention to effectively integrate follow-up and clinical data.} 
    \label{overview}
\end{figure*}
\section{Methods}
\label{sec:format}
In this work, we aim to predict the malignancy of pulmonary nodules by exploring the relationships between follow-up imaging data and clinical information, framing it as a binary classification task. Specifically, we introduce the CSF-Net, which consists of three key modules: the spatial feature extraction module, the temporal residual fusion module, and the Cross-Modal Attention Fusion (CMAF) \cite{luo2022cmafgan} module. Each module addresses specific aspects of the predictive process, as detailed in the following subsections.
\subsection{Spatial Feature Extraction Model}
The spatial feature extraction module integrates the 3D ResNet with the Convolutional Block Attention Module (CBAM) \cite{woo2018cbam} to capture the spatial features of lung nodules in CT images. The former provides a multi-level feature representation, utilizing residual connections to prevent gradient vanishing and explosion, effectively extracting edge, texture, and overall morphology features of the nodules. The latter further enhances feature focus, with its channel attention module adjusting weights across channels using a multi-layer perceptron to emphasize critical information, while the spatial attention module applies global average pooling and max pooling to highlight nodule-relevant spatial regions. The synergy between 3D ResNet and CBAM improves the recognition of complex nodule structures, providing rich, accurate feature representations for subsequent diagnostic and classification tasks.
\subsection{Temporal Residual Fusion Module}
\begin{figure}[!h]
\vspace{-0.4em}
    \begin{center}
    \includegraphics[width=8.6cm]{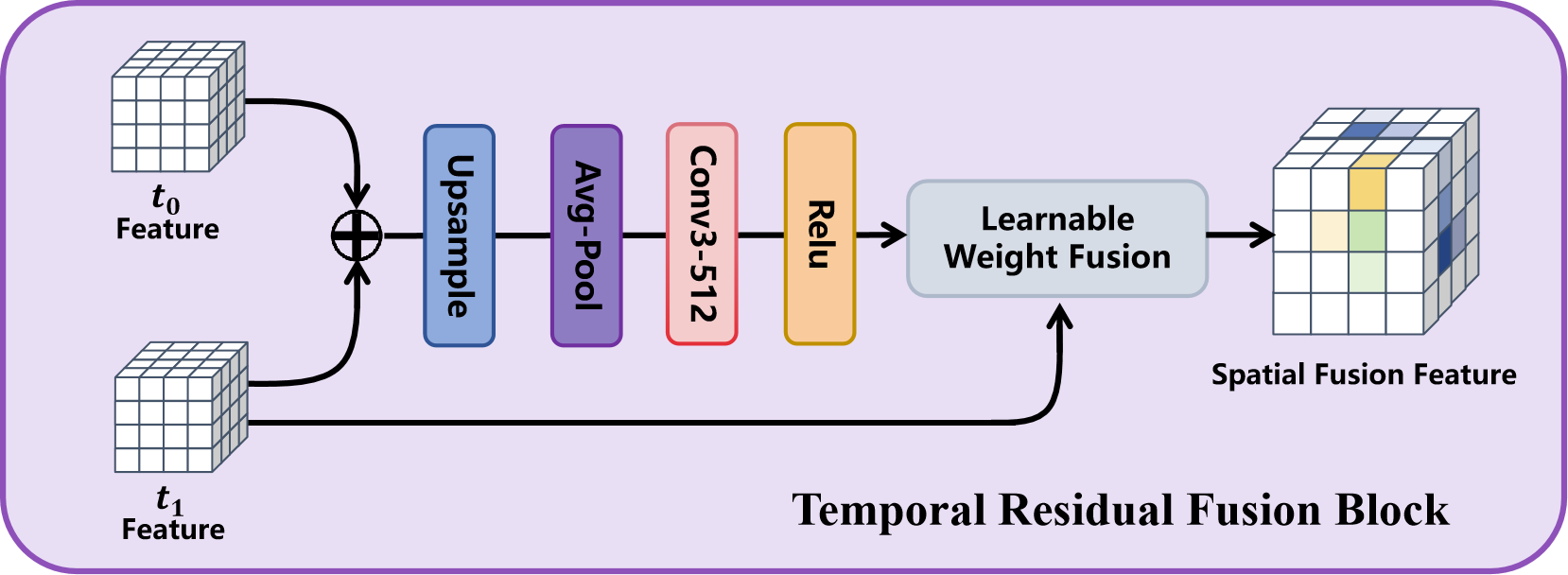}
    \end{center}
   \caption{The architecture of the proposed TRF, designed for deep integration of temporal features, effectively captures changes along the time dimension. }
    \label{tepfusion} 
\end{figure}
To facilitate deep fusion of image features from different time points, we developed a Temporal Residual Fusion (TRF) module. This module inputs features \( t_0 \) and \( t_1 \) from distinct periods and employs a multi-level feature fusion strategy to explore the complementarity of multidimensional features. In the high-dimensional feature space, the model captures richer contextual information and complex nonlinear relationships through concatenation operations. The processed features are then upsampled to restore resolution and consolidated using global average pooling. Subsequently, convolutional layers refine multiscale features, enhancing the model's ability to represent complex data patterns, as shown in the equation:
\vspace{0.4em}
\setlength{\abovedisplayskip}{2pt}
\begin{eqnarray}
    \resizebox{0.43\textwidth}{!}{
    $\mathit{F_{feat}} = \mathit{Conv3D}(\mathit{AvgPool}(\mathit{Upsample}(t_0 \oplus t_1)))$}.
\vspace{-0.3em}    
\end{eqnarray}
Finally, a learnable fusion mechanism dynamically adjusts the feature weights $\lambda_0$ and $\lambda_1$ from different time points, where $\sigma$ is the sigmoid function. This mechanism is expressed as:
\vspace{0.4em}
\begin{eqnarray}
    \resizebox{0.32\textwidth}{!}{
    $\mathit{STF_{feat}} = \mathit{\sigma}(\mathit{\lambda_0})*\mathit{F_{feat}}+\mathit{\sigma}(\mathit{\lambda_1})*\mathit{t_1}$}.
\vspace{-0.3em}    
\end{eqnarray}
This ensures a deep fusion of \( t_1 \) with the processed temporal features, enhancing the ability to capture temporal correlations and integrate information across different times. Consequently, this improves the overall prediction performance.
\subsection{Cross-Modal Attention Fusion Module}
To investigate the intrinsic connections between follow-up imaging data and clinical information, we incorporated a cross-modal fusion module into our framework. This module utilizes spatial data from follow-up images via a spatial feature extraction component and temporal data processed by the TRF module. Clinical information is encoded using a text encoder to extract relevant features, which are then integrated in the CMAF module for comprehensive data fusion. In the CMAF module, the inputs include spatiotemporal features \( x_i \) from follow-up images and clinical features \( y_i \). The module operates as follows:
\begin{equation}
    \mathit{\beta_{j,i}} = \frac{\mathit{exp}(\mathit{s_{ij}})}{\sum_{i=1}^\mathit{S}\mathit{exp}(\mathit{s_{ij}})} \text{, where } \mathit{\mathit{s_{ij}}} = \mathit{q_1}(\mathit{x_i})^{T}\mathit{k_2}(\mathit{y_j}),
\end{equation}
\begin{equation}
    \mathit{\rho_{j,i}} = \frac{\mathit{exp}(\mathit{t_{ij}})}{\sum_{i=1}^\mathit{S}\mathit{exp}(\mathit{t_{ij}})} \text{, where } \mathit{\mathit{t_{ij}}} = \mathit{q_2}(\mathit{y_i})^{T}\mathit{k_1}(\mathit{x_j}),
\end{equation}
\vspace{0.01em}
where $\mathit{S} = \mathit{W}*\mathit{H}$ represents the spatial domain dimensions of the follow-up images, while \( \beta \) and \( \rho \) denote the matching scores in the spatial domains of the images and clinical text, respectively. These scores, \( \beta_{j,i} \) and \( \rho_{j,i} \), are used to weigh the corresponding feature values, creating a final cross-modal attention map.
Ultimately, these mappings are integrated with the spatiotemporal features \( X_i \) from the follow-up images. The combined data is processed through fully connected layers to predict pulmonary nodule malignancy.

\section{Experimental results}
\label{sec:pagestyle}
\subsection{Data Preparation and Preprocess}
The NLST dataset is a pivotal research initiative by the U.S. National Cancer Institute (NCI) aimed at reducing lung cancer mortality in high-risk populations through early imaging screenings. It includes longitudinal CT scans of participants' lungs and relevant clinical information, such as age, gender, smoking status, and screening results as defined by the NLST, with participants receiving annual screenings for three years or until a cancer diagnosis.
Due to the lack of detailed labels for pulmonary nodule locations in the original data, we selected 443 subjects to create a cross-modal spatiotemporal CT dataset, named NLST-cmst. Using data annotated by professional physicians, the ROI regions of pulmonary nodules were extracted. Each participant in this dataset has undergone at least two CT scans, and the malignancy of pulmonary nodules is determined by a pathological gold standard, specifically based on lung cancer diagnoses during follow-up. This approach ensures the data's scientific rigor and reliability.
Our rigorous selection process provides valuable clinical insights for researching the malignant progression of pulmonary nodules, enhancing early detection and treatment strategies.
\subsection{Experimental Details} 
We implemented CSF-Net using the torch-2.1.0-cu12.1-cudnn8.9 framework on a GeForce RTX 3090Ti GPU. In pre-training, we optimized the model with cross-entropy loss and the Adam optimizer, starting with a learning rate of 0.0001 for 200 epochs. To enhance convergence speed and stability, we adjusted momentum parameters $\mathit{\beta_1}$ and $\mathit{\beta_2}$ to 0.5 and 0.999, updating parameters every 20 epochs. This strategy tackles challenges from complex data characteristics. The input to CSF-Net is the ROI region of 3D lung nodules with a size of 16×64×64. 
\vspace{-0.3em}
\subsection{Comparative Experiments}
In this experiment, we evaluated our method against six models: SCANs \cite{9195164}, adaptable for single or multi-timepoint diagnosis of benign and malignant conditions. NAS-Lung \cite{jiang2021learning}, leveraging neural architecture and attention for enhanced interpretability and classification. T-LSTM \cite{liu2022study}, which adds temporal interval sensitivity for precise nodule prediction. RadFusion \cite{zhou2021radfusion}, integrating imaging and EHR data for improved outcomes. DeepCAD \cite{aslani2024enhancing}, which employs a multimodal approach to predict the benign or malignant nature of pulmonary nodules using time series data. MCFN \cite{10539123}, which enhances subtype classification by integrating online data through a multi-module guidance system. Since some of the comparison models do not have publicly available code, we implemented and ran these models based on the descriptions in the literature. The results presented are the actual results from our experiments.
\begin{table}[h]
\centering
\caption{Comparisons of CSF-Net and other methods.}
\resizebox{8.6cm}{!}{
\normalsize
\setlength{\tabcolsep}{1pt} 
\renewcommand{\arraystretch}{1.1}
\begin{tabular}{clllll}
\hline
Method & \textit{Acc} ($\uparrow$) & \textit{Prec} ($\uparrow$) & \textit{F1} ($\uparrow$) & \textit{AUC} ($\uparrow$) & \textit{Rec} ($\uparrow$) \\\hline
SCANs \cite{9195164} & 0.7865 & 0.7667 & 0.7077 & 0.7725 & 0.6571\\
NAS-Lung \cite{jiang2021learning} & 0.8539 & \textcolor{red}{0.8235} & 0.8116 & 0.8910 & 0.8000 \\
T-LSTM \cite{liu2022study} & 0.7645 & 0.7012 & 0.6527 & 0.7778 & 0.6000 \\
DeepCAD \cite{aslani2024enhancing} & 0.8590 & 0.7879 & 0.8254 & 0.8990 & 0.8667 \\
MFCN \cite{10539123} & 0.7949 & 0.7059 & 0.7500 & 0.8903 & 0.8000 \\
RadFusion \cite{zhou2021radfusion} & 0.7753 & 0.8026 & 0.6667 & 0.7693 & 0.6000 \\
\hline 
\textbf{CSF-Net (ours)} & \textcolor{red}{\textbf{$0.8974_{\textcolor{green}{\uparrow0.0384}}$}} & \textcolor{red}{\textbf{$0.8235_{\textcolor{green}{\uparrow0.0000}}$}} & \textcolor{red}{\textbf{$0.8750_{\textcolor{green}{\uparrow0.0399}}$}} & \textcolor{red}{\textbf{$0.9389_{\textcolor{green}{\uparrow0.0496}}$}} & \textcolor{red}{\textbf{$0.9333_{\textcolor{green}{\uparrow0.0666}}$}} \\ \hline
\end{tabular}}
\label{tab:comparision}
\end{table} 

To ensure a fair comparison, we evaluated our proposed model against the six aforementioned models using five key metrics: Accuracy (\textit{Acc}), Precision (\textit{Prec}), Area Under the Curve (\textit{AUC}), F1-score (\textit{F1}), and Recall (\textit{Rec}). These metrics provide a comprehensive assessment of the models' predictive accuracy, precision, and overall performance. As illustrated in Table \ref{tab:comparision}, our proposed model exhibits superior performance across all metrics, except for \textit{Prec} which ties with NAS-Lung. Our model demonstrates notable improvements, increasing \textit{Acc}, \textit{AUC}, \textit{F1}, and \textit{Rec} by at least 0.0384, 0.0496, 0.0399, and 0.0666 over other models. Among the evaluated models, only NAS-Lung matched our model in \textit{Prec}. Except for \textit{Prec}, the model DeepCAD, which also uses multi-modal and multi-temporal fusion, achieves the best performance. This also implies that our model, which uses the multimodal and multi-temporal fusion strategy, is superior to other models that only use a single modality or a single time.

\subsection{Ablation Study}
We conducted an ablation study to assess how various components of our model contribute to its performance. The results are summarized in Table \ref{tab:Ablation}. We employed the same five metrics to evaluate model performance under various configurations. By comparing the full model to versions with certain components removed, we could observe the impact of each module on the overall performance.
\begin{table}[h]
\vspace{-0.5em}
\centering
\caption{Comparison of modal ablation experiment.}
\resizebox{8.6cm}{!}{
\normalsize
\begin{tabular}{cccccc}
\hline
Method & \textit{Acc} ($\uparrow$) & \textit{Prec} ($\uparrow$) & \textit{F1} ($\uparrow$) & \textit{AUC} ($\uparrow$) & \textit{Rec} ($\uparrow$) \\\hline
$\mathit{t_0}$ image & 0.7079	&0.6957	&0.5517	&0.6688	&0.6688 \\
$\mathit{t_0}$ image+clinical & 0.7308	&0.6552	&0.6441	&0.7278	&0.6333 \\
$\mathit{t_1}$ image & 0.8462	&0.7500 & 0.8182 & 0.9021	& 0.9000  \\
$\mathit{t_1}$ image+clinical & 0.8590 & 0.8065 &  0.8197 & 0.8792 & 0.8333  \\
without CMAF+clinical & 0.8718	&0.7941	&0.8438	&0.9149	&0.9000 \\
without temporal fusion & 0.8718	&0.7778	&0.8485	&0.9028	&0.9333 \\
\hline 
\textbf{CSF-Net (ours)} & \textbf{0.8974} & \textbf{0.8235} & \textbf{0.8750} & \textbf{0.9389} & \textbf{0.9333} \\ \hline
\end{tabular}}
\label{tab:Ablation}
\vspace{-1em}
\end{table}

\textbf{1) Impact of Using \( \boldsymbol{t_0} \) vs. \( \boldsymbol{t_1} \) Images Alone:} When only \( t_0 \) images were used, model performance dropped significantly, with \textit{Acc} falling from 0.8974 to 0.7079, along with decreases in \textit{Prec} and \textit{Rec}. However, using only \( t_1 \) images resulted in considerable performance improvements, with increased \textit{Acc}, \textit{Prec}, \textit{Rec}, and \textit{F1} by 0.1666, 0.1683, 0.3000, and 0.2309, respectively. The difference indicates \( t_1 \) images capture more pronounced features of disease progression, while \( t_0 \) images alone do not fully reflect lesion evolution. 

\textbf{2) Impact of Clinical Data to \( \boldsymbol{t_0} \) and \( \boldsymbol{t_1} \) Images:} Introducing clinical data significantly enhanced model performance, especially when combined with \( t_1 \) images, where \textit{Acc}, \textit{Prec}, \textit{Rec}, and \textit{F1} further increased by 0.1283, 0.2313, 0.3167, and 0.1756, respectively. This highlights the critical role of clinical data in complementing image information, providing the model with a more comprehensive understanding of lesion characteristics and thus improving the accuracy and reliability of lesion detection.

\textbf{3) Impact of Cross-Attention and Clinical Data:} Without cross-attention and clinical data, the model’s ability to integrate multimodal information is restricted, resulting in an \textit{Acc} of 0.8718 and a \textit{Prec} of 0.7941. The absence of cross-attention significantly limits the model’s capacity to capture complex relationships between data sources, leading to diminished performance compared to the complete model. Therefore, cross-attention is critical for effectively combining multi-source data, and its absence poses challenges in modeling interactions between various modalities, impacting the model’s overall recognition performance.

\textbf{4) Impact of Temporal Fusion:} Eliminating the temporal fusion strategy resulted in a model \textit{Acc} of 0.8718 and a \textit{Prec} of 0.7778. This outcome highlights the critical role of temporal fusion in optimizing the integration of temporal information. Without this component, the model's capability to reconstruct temporal features is impaired, reducing its sensitivity to changes in lesions. This demonstrates that temporal fusion effectively captures feature variations across different time points, and its absence limits the model's ability to leverage temporal information fully. Consequently, this leads to a decline in overall recognition performance.


\section{Conclusions}
Inefficient temporal feature integration and lack of clinical data consideration can reduce the accuracy of pulmonary nodule malignancy predictions. To address this, we propose CSF-Net, a cross-modal spatiotemporal fusion network model for malignancy prediction. CSF-Net includes a TRF module to capture temporal correlations in follow-up data, boosting predictive performance. Experimental results show CSF-Net outperforms other models across multiple metrics, confirming its effectiveness. However, the model currently relies on pre-identified nodule locations. Future work will focus on evolving CSF-Net into an end-to-end model that directly detects and predicts malignancy from CT images, enhancing clinical utility.

\section{Compliance with ethical standards}
\label{sec:ethics}
This study retrospectively utilized human subject data obtained from the publicly accessible National Lung Screening Trial dataset. The license accompanying this open-access data confirms that no ethical approval is required.

\section{Acknowledgments}
\label{sec:acknowledgments}
This work was supported by the Open Project Program of the State Key
Laboratory of CAD\&CG, Zhejiang University (No.A2410), Zhejiang Provincial Natural Science Foundation of China (No.LY21F020017), National Natural Science Foundation of China (No.61702146, 62076084), GuangDong Basic and Applied Basic Research Foundation
(No.2022A1515110570), Shenzhen Longgang District Science and Technology Innovation Special Fund (No. LGKCYLWS2023018), and Shenzhen Science and Technology Program (No.KCXFZ20201221173008022).

\renewcommand{\baselinestretch}{0.3}  
\bibliographystyle{IEEEbib}
\bibliography{refs}

\begin{thebibliography}{10}

\bibitem{bray2024global}
Freddie Bray, Mathieu Laversanne, Hyuna Sung, Jacques Ferlay, Rebecca~L Siegel, Isabelle Soerjomataram, and Ahmedin Jemal,
\newblock ``Global cancer statistics 2022: Globocan estimates of incidence and mortality worldwide for 36 cancers in 185 countries,''
\newblock {\em CA: A Cancer Journal for Clinicians}, vol. 74, no. 3, pp. 229--263, 2024.

\bibitem{national2011national}
National Lung Screening Trial~Research Team,
\newblock ``The national lung screening trial: overview and study design,''
\newblock {\em Radiology}, vol. 258, no. 1, pp. 243--253, 2011.

\bibitem{jiang2021learning}
Hanliang Jiang, Fuhao Shen, Fei Gao, and Weidong Han,
\newblock ``Learning efficient, explainable and discriminative representations for pulmonary nodules classification,''
\newblock {\em Pattern Recognition}, vol. 113, pp. 107825, 2021.

\bibitem{8642524}
Fangzhou Liao, Ming Liang, Zhe Li, Xiaolin Hu, and Sen Song,
\newblock ``Evaluate the malignancy of pulmonary nodules using the 3-d deep leaky noisy-or network,''
\newblock {\em IEEE Transactions on Neural Networks and Learning Systems}, vol. 30, no. 11, pp. 3484--3495, 2019.

\bibitem{liu2022study}
Xindong Liu, Mengnan Wang, and Rukhma Aftab,
\newblock ``Study on the prediction method of long-term benign and malignant pulmonary lesions based on lstm,''
\newblock {\em Frontiers in Bioengineering and Biotechnology}, vol. 10, pp. 791424, 2022.

\bibitem{9195164}
Benjamin~P. Veasey, Justin Broadhead, Michael Dahle, Albert Seow, and Amir~A. Amini,
\newblock ``Lung nodule malignancy prediction from longitudinal ct scans with siamese convolutional attention networks,''
\newblock {\em IEEE Open Journal of Engineering in Medicine and Biology}, vol. 1, pp. 257--264, 2020.

\bibitem{khademi2023spatio}
Sadaf Khademi, Shahin Heidarian, Parnian Afshar, Farnoosh Naderkhani, Anastasia Oikonomou, Konstantinos~N Plataniotis, and Arash Mohammadi,
\newblock ``Spatio-temporal hybrid fusion of cae and swin transformers for lung cancer malignancy prediction,''
\newblock in {\em ICASSP 2023-2023 IEEE International Conference on Acoustics, Speech and Signal Processing (ICASSP)}. IEEE, 2023, pp. 1--5.

\bibitem{10635747}
Zhaoxin Guo, Zhipeng Wang, Ruiquan Ge, Jianxun Yu, Feiwei Qin, Yuan Tian, Yuqing Peng, Yonghong Li, and Changmiao Wang,
\newblock ``Pe-mvcnet: Multi-view and cross-modal fusion network for pulmonary embolism prediction,''
\newblock in {\em 2024 IEEE International Symposium on Biomedical Imaging (ISBI)}, 2024, pp. 1--5.

\bibitem{yu2025prnet}
Xinlei Yu, Ahmed Elazab, Ruiquan Ge, Jichao Zhu, Lingyan Zhang, Gangyong Jia, Qing Wu, Xiang Wan, Lihua Li, and Changmiao Wang,
\newblock ``Ich-prnet: A cross-modal intracerebral haemorrhage prognostic prediction method using joint-attention interaction mechanism,''
\newblock {\em Neural Networks}, p. 107096, 2025.

\bibitem{10539123}
Saisai Ding, Juncheng Li, Jun Wang, Shihui Ying, and Jun Shi,
\newblock ``Multimodal co-attention fusion network with online data augmentation for cancer subtype classification,''
\newblock {\em IEEE Transactions on Medical Imaging}, pp. 1--1, 2024.

\bibitem{subramanian2020multimodal}
Vaishnavi Subramanian, Minh~N Do, and Tanveer Syeda-Mahmood,
\newblock ``Multimodal fusion of imaging and genomics for lung cancer recurrence prediction,''
\newblock in {\em 2020 IEEE 17th International Symposium on Biomedical Imaging (ISBI)}. IEEE, 2020, pp. 804--808.

\bibitem{ellen2023autoencoder}
Jacob~G Ellen, Etai Jacob, Nikos Nikolaou, and Natasha Markuzon,
\newblock ``Autoencoder-based multimodal prediction of non-small cell lung cancer survival,''
\newblock {\em Scientific Reports}, vol. 13, no. 1, pp. 15761, 2023.

\bibitem{aslani2024enhancing}
Shahab Aslani, Pavan Alluri, Eyjolfur Gudmundsson, Edward Chandy, John McCabe, Anand Devaraj, Carolyn Horst, Sam~M Janes, Rahul Chakkara, Daniel~C Alexander, et~al.,
\newblock ``Enhancing cancer prediction in challenging screen-detected incident lung nodules using time-series deep learning,''
\newblock {\em Computerized Medical Imaging and Graphics}, vol. 116, pp. 102399, 2024.

\bibitem{luo2022cmafgan}
Xiaodong Luo, Xiang Chen, Xiaohai He, Linbo Qing, and Xinyue Tan,
\newblock ``Cmafgan: A cross-modal attention fusion based generative adversarial network for attribute word-to-face synthesis,''
\newblock {\em Knowledge-Based Systems}, vol. 255, pp. 109750, 2022.

\bibitem{woo2018cbam}
Sanghyun Woo, Jongchan Park, Joon-Young Lee, and In~So Kweon,
\newblock ``Cbam: Convolutional block attention module,''
\newblock in {\em Proceedings of the European conference on computer vision (ECCV)}, 2018, pp. 3--19.

\bibitem{zhou2021radfusion}
Yuyin Zhou, Shih-Cheng Huang, Jason~Alan Fries, Alaa Youssef, Timothy~J Amrhein, Marcello Chang, Imon Banerjee, Daniel Rubin, Lei Xing, Nigam Shah, et~al.,
\newblock ``Radfusion: Benchmarking performance and fairness for multimodal pulmonary embolism detection from ct and ehr,''
\newblock {\em arXiv preprint arXiv:2111.11665}, 2021.

\end{thebibliography}

\end{document}